# The formation of close binary systems

Ian A. Bonnell and Matthew R. Bate
*Institute of Astronomy, Madingley Road, Cambridge, CB3 0HA*

**ABSTRACT**

A viable solution to the origin of close binary systems, unaccounted for in recent theories, is presented. Fragmentation, occurring at the end of the secondary collapse phase (during which molecular hydrogen is dissociating), can form binary systems with separations less than 1 au.

Two fragmentation modes are found to occur after the collapse is halted. The first consists of the fragmentation of a protostellar disc due to rotational instabilities in a protostellar core, involving both an $m = 1$ and an $m = 2$ mode. This fragmentation mechanism is found to be insensitive to the initial density distribution: it can occur in both centrally condensed and uniform initial conditions. The second fragmentation mode involves the formation of a rapidly rotating core at the end of the collapse phase which is unstable to the axisymmetric perturbations. This core bounces into a ring which quickly fragments into several components.

The binary systems thus formed contain less than 1 per cent of a solar mass and therefore will need to accrete most of their final mass if they are to form a binary star system. Their orbital properties will thus be determined by the properties of the accreted matter.

## 1 INTRODUCTION

Although significant progress in our understanding of the formation of binary stars has been made recently, the origin of those systems with separations less than 1 au remains, as yet, unclear. None the less, observations have shown that a significant number of main-sequence (Duquennoy & Mayor 1991; Griffin 1992) and pre-main-sequence (Mathieu, Walter & Myers 1989; Mathieu 1992) stars have close companions. At one time, fission of a rapidly rotating protostar was thought to be the most likely mechanism for the formation of close binary systems, but numerical simulations have repeatedly shown that fission does not occur (e.g. Durisen et al. 1986). Instead, the gravitational torques from the bar instability, and ensuing spiral arms, are able to transport the excess angular momentum outwards, leaving a single object and a surrounding disc (see also Bonnell 1994a). Since then, fragmentation has been advanced as one of the most probable mechanisms for the origin of most binary systems (Boss 1988). Although much effort has gone into studying the fragmentation of a collapsing molecular cloud and the ensuing formation of wide (a few au to 0.1 pc) binary systems (Boss 1992a, 1992b, 1993; Bonnell et al. 1991; Bonnell & Bastien 1992; Nelson & Papaloizou 1993), little attention has been paid to the close systems.

Most fragmentation calculations have considered only the isothermal collapse phase (see Bonnell 1994b and references therein). Fragmentation at this stage is unable to form close binary systems because of the densities (and hence the size, $\approx 4$ au [Larson 1969]) at which molecular hydrogen becomes optically thick and fragmentation is suppressed (Boss 1986). It is therefore unlikely that fragmentation during the isothermal collapse phase can form binary systems with separations less than 1 au without significant orbital decay (e.g. Clarke & Pringle 1991).

Another possibility for the formation of close binary systems occurs when the molecular hydrogen starts to dissociate, allowing for a second collapse phase (e.g. Bonnell 1994b). Boss (1989) investigated the possibility of fragmentation during this collapse phase. He found that, although a small $m = 2$ initial density perturbation grew once $\beta$, the ratio of the rotational energy to the absolute value of the gravitational energy, was large (i.e. $\beta \approx 0.3$), the two components were forced to merge due to the growth of spiral arms which transported angular momentum away from the protobinary. An alternative to direct fragmentation during the collapse phase is the fragmentation of a rotationally supported protostellar disc at the end of a collapse phase. The interplay of an $m = 2$ bar-shaped central object with a surrounding disc and continued infall can fragment the disc and thus form a binary system. This process has been found to occur at the end of the isothermal collapse phase (Bonnell 1994a). In this paper, we investigate whether this process can also occur at the end of the second collapse phase, taking into account two additional constraints. First, the initial conditions are that of a pressure (and possibly partially rotationally) supported core with some degree of central condensation, and thus should in general not be too far from virial equilibrium. Secondly, the equation of state in the second collapse phase is due to the dissociation of molecular hydrogen instead of an optically thin gas.

The calculations are presented in Section 2 and the initial conditions are discussed in Section 3. Section 4 presents the results while Section 5 discusses the further evolution of the systems. Our conclusions are included in Section 6.

## 2   CALCULATIONS

The simulations reported here were performed with a three-dimensional Smooth Particle Hydrodynamics (SPH) code (Benz 1990; Benz et al. 1990). The code uses a tree to calculate the gravitational forces and to find the nearest neighbours. It also uses individual time steps for each particle that are limited by the Courant condition. To ensure that no spurious transport of angular momentum occurred due to the artificial viscosity, many tests were performed with different values of the artificial viscosity parameters and with a switch to reduce the artificial viscosity in shear flows (Benz 1990). No transport of angular momentum occurred before the development of the non-axisymmetric perturbations. The transport of angular momentum once non-axisymmetric perturbations are present can be directly attributed to gravitational torques (Bonnell 1994a). The only effect of the different levels of the artificial viscosity was to increase or decrease the energetics of the bounce that occurs at the end of the second collapse phase (note that a bounce will occur at the end of any dynamical collapse phase). The different levels of the bounce affect exactly which fragmentation mode occurs (e.g. a more violent bounce is more likely to undergo ring fragmentation, see Section 4.2), but the fragmentation itself (see Section 4), occurring after the bounce, was not affected.

As in Bonnell (1994a), a polytropic equation of state is used, but with values of $\gamma$ that apply to the second collapse phase:

$$P = K\rho^\gamma$$

with $\gamma$ depending on the density $\rho$ such that

$\gamma = \gamma_1 = 7/5 \qquad \rho < \rho_{c1}$

$\gamma = \gamma_2 = 1.1 \qquad \rho_{c1} < \rho < \rho_{c2}$

and

$\gamma = \gamma_3 = 5/3 \qquad \rho > \rho_{c2}.$

The values of $\gamma$ are taken from Tohline (1982) and correspond to where molecular hydrogen is optically thick ($\gamma_1 = 7/5$), where molecular hydrogen is dissociating ($\gamma_2 = 1.1$) and where the hydrogen is in atomic form ($\gamma_3 = 5/3$). The critical values $\rho_{c1}$ and $\rho_{c2}$ are chosen to be $5.66 \times 10^{-8}$ g cm$^{-3}$ and $1.0 \times 10^{-2}$ g cm$^{-3}$, respectively, where the former is chosen as being where the temperature reaches 2000 K assuming that it starts at a temperature of 10 K at $\rho = 1.0 \times 10^{-13}$ g cm$^{-3}$. Lower values of $\rho_{c2}$ (e.g. $10^{-3}$ and $10^{-4}$ g cm$^{-3}$) were tried to ensure that the exact value when collapse was halted does not affect the results. The only effect of lower values of $\rho_{c2}$ is a slight change in the scale at which fragmentation occurs, and hence the separation of the resultant binary.

### 2.1   Effects of the equation of state

Although a polytropic equation of state is a crude representation of the real equation of state, it allows us to follow the dynamics of the collapse phase. The exact softening due to the dissociation of molecular hydrogen should not greatly change the results as long as $\gamma_2 \approx 1.1$. The equation of state used by Boss (1989) is actually softer than the one used here during H$_2$ dissociation. The approximate form of the equation of state does imply that the calculations correspond to either a higher value of $J_0$ (the ratio of the absolute value of the gravitational energy to the thermal energy) if $\gamma_2 > 1.1$, or a lower value of $J_0$ if $\gamma_2 < 1.1$ (as is the case for the equation of state used by Boss [1989]).

The fragmentation modes presented below depend on the rotational instability of the final core. Rotational instabilities are expected if $\beta_f$, the final value of $\beta$, is greater than a certain critical value ($\beta_f \gtrsim 0.3$; e.g. Tassoul 1978). Here we investigate how the exact value of $\gamma_2$ can affect $\beta_f$ and hence the probability of fragmentation. For larger values of $\gamma_2$, there is a possibility of finding an equilibrium configuration and thus halting the collapse before $\rho = \rho_{c2}$ (Tohline 1984). In an isothermal (or nearly isothermal) collapse, the thermal energy does not increase appreciably, and an equilibrium can therefore only be reached if $\beta_f \approx 0.5$. Alternatively, in a non-isothermal collapse, the thermal energy can increase sufficiently to be an important source of support. In this case, $\beta_f$ can be relatively small, and below the limit for dynamical instability (i.e. $\beta_f < 0.3$). An extreme example of this is that, if $\gamma_2 > 4/3$, the thermal energy can halt collapse without any rotational support ($\beta_f = 0$). Following Tohline (1984), and using Maclaurin spheroids to judge the endpoint of a collapse, we can determine for what phase space of the initial conditions $(J_0, \beta_0)$ the core will collapse all the way to $\rho_{c2}$. For example, if $\gamma_2 = 1.2$, collapse can be halted before reaching $\rho_{c2}$ if $\beta_0 > 0.1$. Furthermore, for those cores that collapse completely to $\rho_{c2}$, as long as $\beta_0 > 0.02$, they will have $\beta_f > 0.3$.

To verify this, several simulations were run with $\gamma_2 = 1.2$ during molecular hydrogen dissociation. In general, these simulations evolve in a qualitatively similar way to those with $\gamma_2 = 1.1$. They collapse down to $\rho_{c2}$ and have a $\beta_f > 0.3$. An exception occurs for some simulations that started from near virial equilibrium, but with significant rotation ($\beta_0 \geq 0.1$). These last simulations reached an equilibrium at densities $\rho < \rho_{c2}$, with $\beta_f < 0.3$, and did not undergo any rotationally driven instabilities (see also Tohline 1984).

## 3   INITIAL CONDITIONS

Following Boss (1989), we simulate the second collapse phase from 'initial' conditions representing the outer core (formed from the isothermal collapse phase) at the time when the second collapse phase occurs. The envelope that is still infalling on to the outer core is ignored. This does not greatly affect the collapse, since the time-scale for the envelope's infall is much longer than the free-fall time of the core. This does imply, however, that the constant-volume boundary conditions used here are a limitation. As the gas collapses, the density increases in the central regions and correspondingly decreases near the boundary. This causes the pressure to drop at the boundary. In reality, the outer envelope would maintain a near-constant pressure on the collapsing core. A constant-pressure boundary would have the effect of increasing the Jeans number of the core as it would compress the less-dense regions, maintaining a minimum density comparable to the original density at the boundary. This would enhance the possibility of fragmentation.

The initial conditions can be characterized in two ways: by the initial configuration of the outer core that collapses and by how close they are to virial equilibrium. The initial configurations are in two types. The first type has uniform initial density with $R_0 = 5.9 \times 10^{12}$ cm (0.4 au), while the second has a Gaussian density profile:

$$\rho = \rho_o \times \exp\left[-a\left(\frac{R}{R_0}\right)^2\right],$$

where the constant $a$ is picked such that the density at the boundary ($\rho_{R_0}$) is 1/20 of the central density ($\rho_0$), and an initial radius of $R_0 = 11.8 \times 10^{12}$ cm (0.8 au). The main difference between the two is that, for the second case, the initial density is lower than the critical density for the second collapse phase to begin, and this, combined with the slight central condensation, means that the core has to contract quasi-statically until the central region which contains a Jeans mass has $\rho > \rho_{c1}$. This is probably a more realistic initial condition for the second collapse phase. To ensure that the results are not sensitive to the initial density profile, one simulation was run with a $1/r$ density profile. This simulation evolves as do the other simulations that are initially centrally condensed, forming a rotationally unstable core with surrounding disc. Unfortunately, this simulation was not followed long enough to see whether fragmentation would eventually occur.

The simulations can also be divided into two categories by the amount of rotation initially present, and hence how close they are to virial equilibrium. Those of the first type start from conditions close to virial equilibrium,

$$\frac{1}{J_0} + \beta_0 \approx 0.5,$$

and therefore have significant rotational support. Those of the second type have little rotational support, and are not in approximate virial equilibrium (corresponding to the calculations reported in Bonnell [1994a] for the case of the first collapse phase).

The calculations were performed with 10 000 or 20 000 particles. The masses used vary between 0.02 and 0.1 $M_\odot$, giving mean densities that vary between 0.1 and 4 times $\rho_{c1}$. Solid body rotation with $\beta_0$ varying between 0.03 and 0.25 was used to simulate initial conditions with very little or significant rotational support. In terms of fragmentation, solid body rotation should be the worst case since differential rotation has been found to increase fragmentation (Myhill & Kaula 1992).

## 4 RESULTS

In all the simulations, the collapse proceeds axisymmetrically until the bounce (which halts the collapse phase) has occurred. No fragmentation, due to any small initial perturbations, or due to the cloud's rotation, occurs before the bounce. Even simulations that start from high $J_0$ and $\beta_0$ (e.g. $J_0 = 10$, $\beta_0 = 0.1$) do not fragment before the core bounces. Thus the small perturbations inherent in a discretized simulation are not able to grow sufficiently for fragmentation to occur before rotational and pressure support increases sufficiently to halt the collapse.

In the simulations that started from centrally condensed initial conditions, the central density, initially below the critical density ($\rho_{c1}$; where molecular hydrogen starts to dissociate), decreases during the initial stages of the evolution. Hence the central regions do not contain a Jeans mass and therefore have to wait for the contraction of the less-dense regions before collapse can proceed. This implies that the degree of central condensation used in the initial conditions here is sufficient.

The results can be divided into two sections. The first section is for the centrally condensed initial conditions and for cores with uniform initial density but with lower values of $J_0$. These simulations underwent disc fragmentation (see Section 4.1 below). The second set of results involve more unstable initial conditions, such that a greater number of Jeans masses are involved in the collapse. These simulations underwent ring fragmentation (see Section 4.2 below). All of the latter simulations started from an initially uniform density. The main difference between the two sets of results is the number of Jeans masses involved at the bounce that occurs at the end of the second collapse phase.

### 4.1 Disc fragmentation

In these simulations, the protostellar disc fragments due to rotational instabilities. These simulations can be divided into two groups. The first group, where the collapse remains axisymmetric until the bounce occurs, undergo evolutions similar to that previously investigated by Bonnell (1994a) in the case of the first collapse phase. These will only be briefly discussed here. Important differences in the simulations are the value of $\gamma$ during collapse ($\gamma = 1.1$) and the variety of initial conditions (centrally condensed or uniform density; small $\beta_0$ or $\beta_0 \approx 1/2 - 1/J_0$). In all the simulations, the cloud collapses to form a rotationally supported disc with a pressure supported core inside. Rotational instabilities develop in the core and disc, involving both an $m = 1$ and an $m = 2$ non-axisymmetric mode. The $m = 2$ mode grows first, once the core has $\beta > 0.3$. The $m = 1$ mode grows afterwards, being driven by the $m = 2$ mode (see Bonnell 1994a).

The second group differ in that the collapsing core does not remain completely axisymmetric. In these simulations, rotational support is initially important (in contrast to the simulations reported in Bonnell [1994a]), and the $m = 1$ and $m = 2$ modes can grow concurrently during the collapse. This can be understood as $\beta > 0.3$ well before the bounce occurs. The collapse is thus slowed down and the instabilities have time to grow. For a discussion on the growth of low-$m$ non-axisymmetric modes during collapse, the reader is referred to Boss (1982). It is also of interest to note here that Woodward, Tohline & Hachisu (1994) have found that $m = 1$ instabilities can grow in massive protostellar discs for moderate values of $\beta$ ($\beta \approx 0.2$).

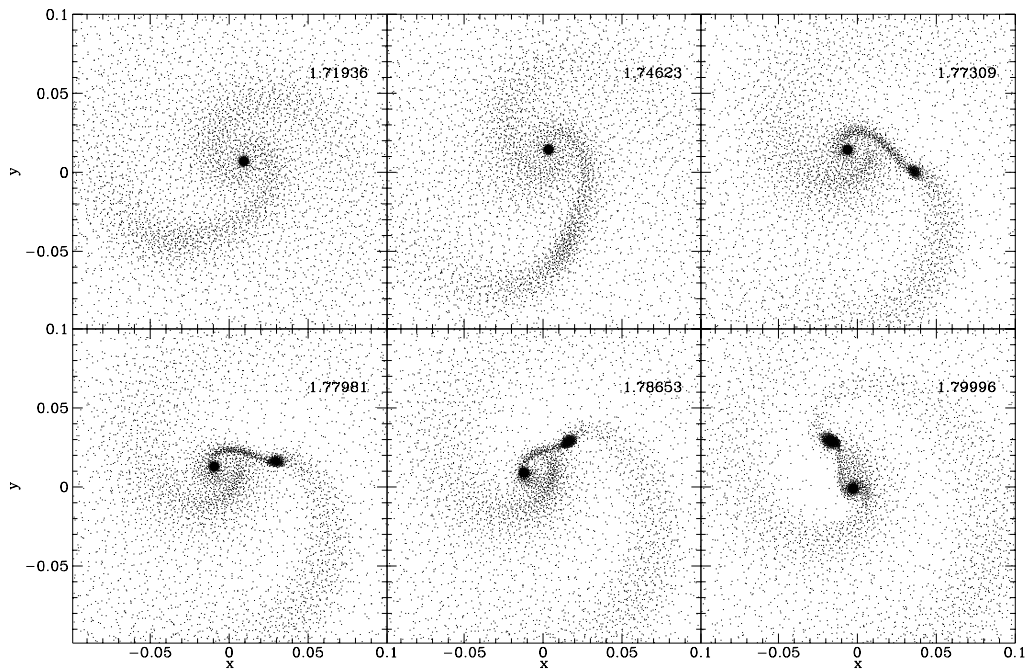

**Figure 1.** The particle positions projected on to the equatorial plane for a simulation involving the fragmentation of a protostellar disc around the inner protostellar core. The simulation was performed with 10 340 particles, $J_0 = 2.5$ and $\beta_0 = 0.1$. The time in units of the free-fall time ($t_{\rm ff} = 9.26 \times 10^6$ s) is given in the upper right hand corner of each panel. The axes are in units of the initial radius ($5.9 \times 10^{12}$ cm).

In both cases, the spiral arms due to these non-axisymmetric ($m = 1$ and $m = 2$) modes sweep through the disc and gather matter together to form a gravitationally unstable condensation. This condensation collapses and forms a companion. Any rotational support that the matter has is overcome by the transfer of angular momentum by the gravitational torques (Bonnell 1994a). This process is illustrated in Fig. 1.

This result does not depend on the initial density profile, as clouds with uniform or centrally condensed initial densities form rotationally unstable cores and discs. Furthermore, the clouds that start with $\rho < \rho_{c1}$ contract quasi-statically and thus lose their initial conditions before collapse proceeds. This can be easily understood since, in any collapse that is non-homologous (i.e. for $1 < \gamma < 4/3$), the density profile will evolve to a $\rho \propto r^{-n}$ (with $2 < n < 3$) form in the outer part of the cloud. Therefore, since the fragmentation occurs after the collapse is halted, the initial density profile should not be important.

### 4.2  Ring fragmentation

These simulations were performed to see if direct fragmentation during collapse is possible. As stated above, none of the simulations fragment before the collapse is halted. In-stead, the collapse proceeds nearly axisymmetrically until rotational support increases sufficiently to stop the collapse (see Bonnell 1994a). Since the cloud has a larger value of $J_0$ than do the simulations discussed above or in Bonnell (1994a), the core contains several Jeans masses when rotational support stops the collapse. Furthermore, the infalling gas compresses the core past the point of rotational support. The core then bounces outwards in the equatorial plane. At the point of maximum compression, if the value of $\beta$ of the rotationally supported core is large (e.g. $\beta > 0.45$), then the core is unstable to axisymmetric perturbations and can bounce into the form of a ring. The ring is itself highly unstable to non-axisymmetric perturbations and quickly fragments (see also Norman & Wilson 1978). This process can form two or more fragments, depending on the number of Jeans masses contained in the core when the bounce occurs. This process is illustrated in Fig. 2.

### 5  FURTHER EVOLUTION

Although the above section shows that fragmentation (via either a disc or a ring) at the end of the second collapse phase is capable of forming close binary systems, it must be noted that these systems contain very little mass. Since the second collapse phase occurs when $\approx 1$ to 10 % of a solar mass is contained in the first (outer) protostellar core, and since the fragmentation process forms objects with about 10 % of the total mass involved, the binary systems thus formed have less than 1 per cent of a solar mass. To form a binary system where the components have stellar masses, these seeds will have to grow through accretion (e.g. Artymowicz 1983). To follow this evolution is beyond the capabilities and scope of the present study. This does imply, however, that the final orbital parameters will depend crucially on the subsequent accretion.

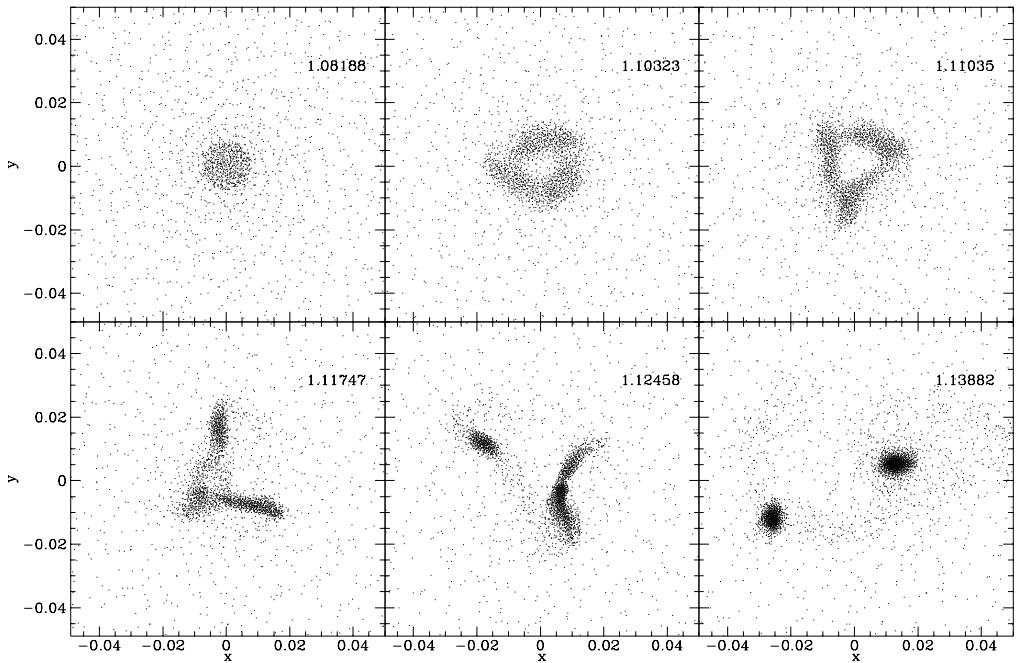

**Figure 2.** The particle positions projected on to the equatorial plane for a simulation involving ring fragmentation. The simulation was performed with 10 340 particles, $J_0 = 2.75$ and $\beta_0 = 0.03$. The time in units of the free-fall time ($t_{\rm ff} = 8.74 \times 10^6$ s) is given in the upper right hand corner of each panel. The axes are in units of the initial radius ($5.9 \times 10^{12}$ cm).

Furthermore, a competing process involving angular momentum loss could force the two components to merge together long before the accretion has finished. One such process that has been investigated is the dynamical interaction of a binary system with a non-self-gravitating circumbinary disc (Artymowicz et al. 1991). It was found that the loss of angular momentum through the gravitational torques operating between the binary and the disc can cause significant orbital evolution, decreasing the separation, especially at periastron. This process is liable to play an important role in the development of a protobinary system where most of the matter that has to be accreted will have higher angular momentum and thus form a circumbinary disc.

Alternatively, the embedded binary can force the circumbinary disc to fragment (Bonnell & Bate 1994). The fragmentation of the circumbinary disc occurs in a similar manner to the fragmentation process discussed here. This will ensure the survival of a close binary system even if the protobinary does merge. This binary decay and circumbinary disc fragmentation process could be repetitive throughout the long accretion phase, resulting in the formation of a close binary star at the end of the accretion phase.

## 6 CONCLUSIONS

The fragmentation and formation of binary systems during, and after, the second collapse phase are investigated. Fragmentation does not occur during the collapse phase itself but can occur via two mechanisms once the collapse is halted. First, fragmentation of a protostellar disc can occur, induced by the combination of $m = 1$ and $m = 2$ non-axisymmetric modes. These modes combine to gather sufficient matter together in a condensation to be gravitationally unstable and collapse. This result, extending a previous study by Bonnell (1994a) of the (first) isothermal collapse phase, is unaffected by differences in the initial conditions (e.g. centrally condensed and near virial equilibrium) and the exact equation of state. The exact initial density distribution does not affect this mechanism, as the fragmentation occurs at the end of the collapse phase, by which point all memory of the initial density profile is erased by the collapse.

The second fragmentation mechanism involves the development of an axisymmetric ring perturbation during the bounce that occurs at the end of the second collapse phase. If the collapse is halted by rotational support, the central regions contain a number of Jeans masses, and the collapse involves significant radial motion such that it is halted only when compressed beyond equilibrium, it can be unstable to the development of axisymmetric perturbations. It then bounces into a ring which can fragment into several pieces. The binary and multiple systems formed from these processes are close systems with separations of a few $R_\odot$.

The binary and multiple systems thus formed contain very little mass. This is due to the fact that the second collapse phase occurs when only a few per cent of a solar mass is involved. The formation of a binary system with stellar masses (and the binary's properties) from these seeds will therefore depend critically on the accretion process.

## ACKNOWLEDGMENTS

We would like to thank Jim Pringle, Cathie Clarke, Andrea Ghez, Alan Boss and Christian Boily for many useful comments and a critical reading of the manuscript. We would also like to thank the IoA system managers for their continuing efforts to provide the desired computational resources. IAB is grateful to PPARC for a postdoctoral fellowship. MRB is grateful for a scholarship from the Cambridge Commonwealth Trust.

## REFERENCES

Artymowicz P., 1983, Acta Astron., 33, 223
Artymowicz P., Clarke C. J., Lubow S. H., Pringle J. E., 1991, ApJ, 370, L35
Benz W., 1990, in Buchler J.R., ed., The Numerical Modeling of Nonlinear Stellar Pulsations. Kluwer, Dordrecht, p. 269
Benz W., Bowers R. L., Cameron A. G. W., Press W., 1990, ApJ, 348, 647
Bonnell I. A., 1994a, MNRAS, 269, 837
Bonnell I. A., 1994b, in Clemens D. P., Barvainis R., eds, ASP Conf. Ser. Vol. 65, Clouds, Cores and Low Mass Stars. Astron. Soc. Pac., San Francisco, p. 115
Bonnell I., Bastien P., 1992, ApJ, 401, 654
Bonnell I. A., Bate M. R., 1994, MNRAS, 269, L45
Bonnell I., Martel H., Bastien P., Arcoragi J.-P., Benz W., 1991, ApJ, 377, 553
Boss A. P., 1982, ApJ, 159, 165
Boss A. P., 1986, ApJS, 62, 519
Boss A. P., 1988, Comments in Astrophys. 12, 169
Boss A. P., 1989, ApJ, 346, 336
Boss A. P., 1992a, in Sahade J., McCluskey G., Kondo Y., eds, Close Binaries. Kluwer, Dordrecht, p. 355
Boss A. P., 1992b, in Hartkopf W., McAlister H., eds, Proc. IAU Colloq. 135, Complementary Approaches to Double and Multiple Star Research. Astron. Soc. Pac., San Francisco, p. 195
Boss A. P., 1993, ApJ, 410, 157
Clarke C. J., Pringle J. E., 1991, MNRAS, 249, 588
Duquennoy A., Mayor M., 1991, A&A, 248, 485
Durisen R. H., Gingold R. A., Tohline J. E., Boss A. P., 1986, ApJ, 305, 281
Griffin R. F., 1992, in Hartkopf W., McAlister H., eds, Proc. IAU Colloq. 135, Complementary Approaches to Double and Multiple Star Research. Astron. Soc. Pac., San Francisco, p. 98
Larson R. B., 1969, MNRAS, 145, 271
Mathieu R. D., 1992, in Kondo Y., Sistero R. F., Polidan R. S., eds, Evolutionary Processes in Interacting Binary Stars. Kluwer, Dordrecht, p. 21
Mathieu R. D., Walter F. M., Myers P. C., 1989, AJ, 98, 987
Myhill E. A., Kaula W. M., 1992, ApJ, 386, 578
Nelson R., Papaloizou J. C., 1993, MNRAS, 265, 905
Norman M., Wilson J., 1978, ApJ, 224, 497
Tassoul J. L., 1978, Theory of Rotating Stars. Princeton University Press, Princeton, p. 280
Tohline J. E., 1982, Fundam. Cosmic Phys., 8, 1
Tohline J. E., 1984, ApJ, 285, 721
Woodward J. W., Tohline J. E., Hachisu I., 1994, ApJ, 420, 247